\documentstyle{article}
\newcommand{\beq}{\begin{equation}}
\newcommand{\eeq}{\end{equation}}
\newcommand{\beqa}{\begin{eqnarray}}
\newcommand{\eeqa}{\end{eqnarray}}
\newcommand{\bea}{\begin{eqnarray}}
\newcommand{\eea}{\end{eqnarray}}
\newcommand   {\ELJ}     {E_{\mbox{\tiny LJ}}}
\newcommand   {\rgyr}    {r_{\mbox{\tiny gyr}}}
\newcommand   {\Ca}      {C${}^\alpha$\ }

\newcommand   {\bv}      {\hat b}
\newcommand   {\ev}[1]   {\langle #1\rangle}

\input{psfig}
\begin{document}
\begin{titlepage}

\begin{flushright}
LU TP 96-28\\
\today\\
\end{flushright}

\vspace{0.8in}

\LARGE
\begin{center}
{\bf Local Interactions and Protein Folding: \\
A 3D Off-Lattice Approach}\\
\vspace{.3in}
\large
Anders Irb\"ack\footnote{irback@thep.lu.se}, 
Carsten Peterson\footnote{carsten@thep.lu.se}, 
Frank Potthast\footnote{frank@thep.lu.se} and
Ola Sommelius\footnote{ola@thep.lu.se}\\
\vspace{0.10in}
Complex Systems Group, Department of Theoretical Physics\\ 
University of Lund,  S\"{o}lvegatan 14A,  S-223 62 Lund, Sweden \\
\vspace{0.3in}	

Submitted to {\it Journal of Chemical Physics}

\end{center}
\vspace{0.3in}
\normalsize
Abstract:

The thermodynamic behavior of a three-dimensional off-lattice model for 
protein folding is probed. The model has only two types of residues,  
hydrophobic and hydrophilic. In absence of local interactions, native 
structure formation does not occur for the temperatures considered. By 
including sequence independent local interactions, which qualitatively 
reproduce local properties of functional proteins, the dominance of a native 
state for many sequences is observed. As in lattice model approaches, folding 
takes place by gradual compactification, followed by a sequence dependent 
folding transition. Our results differ from lattice approaches in that bimodal 
energy distributions are not observed and that high folding temperatures 
are accompanied by relatively low temperatures for the peak of the specific 
heat. Also, in contrast to earlier studies using lattice models, our results 
convincingly demonstrate that one does not need more than two types of 
residues to generate sequences with good thermodynamic folding properties 
in three dimensions. 

\end{titlepage}

\newpage
\section{Introduction}

In the process of unveiling central issues in the thermodynamics and kinetics 
of protein folding, simplified models where the amino acid residues 
constitute the basic entities seem to exhibit many non-trivial 
and interesting properties \cite{Karplus:95}. In particular lattice 
model approaches with contact interactions only have become increasingly 
popular. The lattice and contact term approximations may seem 
drastic. Nevertheless, it turns out that such  simple models are able to catch 
non-trivial aspects of the folding problem. This is interesting and 
encouraging. However, it does not imply that the approximations involved are 
understood. It is therefore crucial to pursue the study of alternative models,
e.g. exploring off-lattice models with Lennard-Jones interactions.  

The major advantage of lattice models is computational efficiency, at least 
for small chain sizes. With improved algorithms and faster computers this 
advantage is losing in importance. In this paper we propose and study an 
extension to three dimensions (3D) of a two-dimensional (2D) off-lattice 
model~\cite{Stillinger:93} that was successfully studied in 
Refs.~\cite{Irback:95b,Irback:96b}. This model contains only two types of 
amino acids, hydrophobic and hydrophilic, and the key part of the energy 
function is a species-dependent term that favors 
the formation of a hydrophobic core. However, as will be demonstrated below,  
this term alone is not sufficient in order to have thermodynamically dominant 
native states. This observation is reminiscent of existing lattice model 
results~\cite{Shakhnovich:94,Yue:95}, and could be taken as an indication 
that it is essential to have more than two different types of amino acids. 
In this paper we explore another possibility; do the 
folding properties depend on species-independent local interactions sticking 
to two amino acid types only? It should be noted that both lattice and 
2D off-lattice models implicitly contain local interactions. In the lattice case
the mere presence of a discretized space ``stiffens'' the dynamics locally, 
and in two dimensions the continuum movements are hampered by compressing 
one dimension.

The purpose of this work is to construct and numerically study a 3D 
generalization of the 2D model of Refs.~\cite{Irback:95b,Irback:96b}. 
As design criteria for such a model we have:
\begin{itemize}
\item The model should give rise to thermodynamically dominant states -- 
i.e. be biologically plausible from a stability point of view.
\item The local interactions should be chosen to at least qualitatively
reproduce bond and torsional angle distributions and local correlations
found in functional proteins. 
\end{itemize}

We propose a simple form for the local interactions, which are found
to play an important role. Without any local interaction in the model,
the local structure of the chains is much more irregular than
for proteins. It turns out
that one can obtain a local structure qualitatively similar to that of
proteins, by adjusting the strength of the local interactions.
Having chosen the local interactions in this way, we reexamine the
overall thermodynamic behavior. We find that not only have the local
properties improved, but the native states have also become more stable.

Furthermore, we examine the structure formation and the properties of the 
folding transition by studying the temperature dependence of various 
thermodynamic and structural variables, and the distributions of these 
quantities at the folding temperature.
These results can be understood in terms of a gradual collapse to 
compact structures where the (sequence dependent) folding transition occurs. 
The qualitative aspects of this picture confirm what has been found in lattice
model calculations \cite{Karplus:95,Socci:95}. However, it turns out 
that our results differ from lattice approaches in that bimodal 
energy distributions are not observed and that high folding temperatures are 
accompanied by relatively low temperatures for the peak of the specific heat. 

The paper is organized as follows. In Sec.~2 the 3D model is defined and 
in Sec.~3 we extract local properties and correlations for functional proteins 
from the PDB data base~\cite{Bernstein:77}. Monte Carlo (MC) methods and 
measurements are described in Sec.~4. The results and the summary are found in
Sec.~5 and 6 respectively.

\section{The Model}

\begin{figure}[htbp]
\vspace{1.5in}
\begin{center}
\vspace{-105mm}
\mbox{\hspace{0mm}\psfig{figure=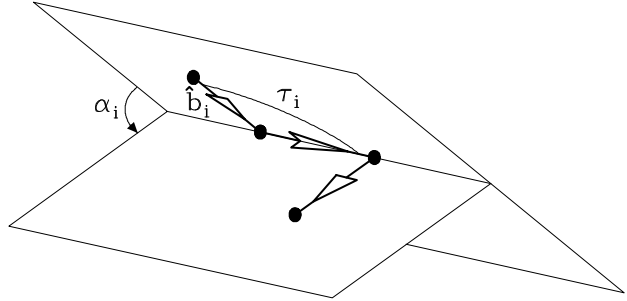,width=15cm,height=20cm}}
\vspace{-85mm}
\end{center}
\caption{Definition of $\bv_i$, $\tau_i$ and $\alpha_i$.}
\label{fig:1}
\end{figure}

We start by defining the simplified geometry of the model. 
Each residue is represented by a single site located at the \Ca position. These sites are 
linked by rigid bonds of unit length, $\bv_i$, to form linear chains living in 
three dimensions. The shape of an $N$-mer is either specified by  the $N-1$ 
bond vectors $\bv_i$ or by $N-2$ bond angles $\tau_i$  and  $N-3$ torsional angles 
$\alpha_i$ (see Fig.~\ref{fig:1}).

The model contains two kinds of residues, $A$ and $B$, which behave
as hydrophobic  ($\sigma_i$=+1) and hydrophilic ($\sigma_i=-1$) 
residues respectively. The energy function is given by
\beq
E(\bv; \sigma) =  -\kappa_1\sum_{i=1}^{N-2}\bv_i\cdot\bv_{i+1}  
- \kappa_2\sum_{i=1}^{N-3}\bv_i\cdot\bv_{i+2} + 
\sum_{i=1}^{N-2}\sum_{j=i+2}^N \ELJ(r_{ij};\sigma_i,\sigma_j)
\label{energy}
\eeq   
where $r_{ij}=r_{ij}(\tau_{i+1},\ldots,\tau_{j-1};
\alpha_{i+1},\ldots,\alpha_{j-2})$ denotes the distance between residues $i$ 
and $j$, and $\sigma_1,\ldots,\sigma_N$ is a binary string that 
specifies the primary sequence. The {\it species-dependent} global interactions
are given by the Lennard-Jones  potential,    
\beq
\ELJ(r_{ij};\sigma_i,\sigma_j)=
4\epsilon(\sigma_i,\sigma_j)\Big( \frac{1}{r_{ij}^{12}}-
\frac{1}{r_{ij}^{6}}\Big)\ .
\label{lj}
\eeq
The depth of the minimum of this potential, $\epsilon(\sigma_i,\sigma_j)$, 
is chosen to favor the formation of a core of $A$ residues
\beq
 \epsilon(\sigma_i,\sigma_j) =
                             \left\{ \begin{array}{ll}
                               {1}            & \mbox{AA}\\
                               &\\
                               {1\over 2} & \mbox{BB, AB}
                              \end{array} \right.\\
\eeq
The two parameters of the energy function, $\kappa_1$ and $\kappa_2$, 
determine the strength of {\it species-independent} local interactions. The model 
will be explored for different values of these two parameters, and our final 
choice will be $\kappa_1=-1$ and $\kappa_2=0.5$.

The behavior of the model at finite temperature $T$ is given  
by the partition function 
\beq
Z(T;\sigma)=\int 
\biggl[\prod_{i=1}^{N-1}d\bv_i\biggr] 
\exp(-E(\bv;\sigma)/T)\ .
\label{z}
\eeq
Let us stress that the interactions in Eq.~\ref{energy} are not chosen so as 
to stabilize the native state of some particular sequence. Rather, our goal is
to study general sequences for a given energy function. We have attempted to 
choose this energy function as simple as possible. Anticipating some 
of results to be presented in Sec.~5, let us here briefly discuss 
the relevance of the different interaction terms. 

{\bf Species-dependent interactions.} It is obvious that the global  
interactions play a key role in the model; these interactions are responsible 
for the compactification of the chain, and for the formation of a hydrophobic 
core. For functional proteins, we find that probability distributions of bond 
and torsional angles depend only weakly on the hydrophobicity pattern for the 
residues involved. In our model, the form of these distributions is very 
sensitive to the choice of global interactions, and it can be strongly 
sequence dependent. In order to avoid this, we have chosen the potentials for 
the three different types of residue pairs to be fairly similar; they are all 
attractive at large separations, and the location of the minimum is the same.

{\bf Species-independent interactions.} It is less clear  how important the local 
interactions are. When studying the behavior of several different sequences in the 
absence of the local interactions ($\kappa_1=\kappa_2=0$),  the stability of the 
native state tends to be very low. Furthermore, we find that local correlations 
along such chains are weak, which is in line with the findings of Ref.~\cite{Socci:94a}. 
For functional proteins the corresponding correlations are fairly strong, which is 
a manifestation of the presence of  secondary structure.  
For these reasons we decided to incorporate the local interactions in the
model. In this way one gets stiffer chains, which implies stronger 
local correlations. In addition, the stability of the native states tends
to improve, as we will see below.

We have studied the model for many different choices of $\kappa_1$ and
$\kappa_2$. The general behavior of the system is fairly insensitive to 
small changes of these parameters. Below we will focus on the results 
obtained for the three choices $(\kappa_1,\kappa_2)$=(0,0),(-1,0) and 
(-1,0.5). 

\section{Local Structures in Functional Proteins}

In a 3D off-lattice model, it is possible to check the local properties against 
those for functional proteins in a direct way. We have probed the local structure in 
two different ways. First we consider the distributions of bond and torsional 
angles. Second we study local correlations along the chain. In this section we 
describe the results one finds for functional proteins. These results will be 
used in Sec.~5 to compare with the qualitative local behavior of our model.

{\bf Distributions of bond and torsional angles.} These distributions for 
functional proteins are well-known~\cite{Levitt:76,Godzik:93} and are included 
here for completeness.  We consider the structure defined by the backbone of 
\Ca atoms. The results were obtained using a set of 505 
selected sequences~\cite{Hobohm:94}\footnote{The May 1996 edition was used.} 
from the Protein Data Bank~\cite{Bernstein:77}. 
This set was obtained by allowing for a maximum of 25\% sequence similarity 
for aligned subsequences of more than 80 residues~\cite{Hobohm:92}. 
Within this set of 505 minimally redundant sequences, 491 contained 
complete backbone information; the others were excluded from our analysis. 
\begin{figure}[tb]
\vspace{1.5in}
\begin{center}
\vspace{-30mm}
\mbox{\hspace{0mm}\psfig{figure=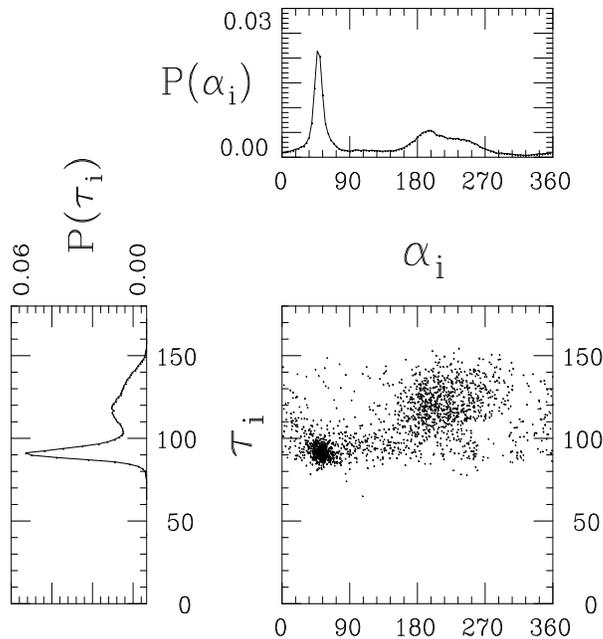,width=7.5cm,height=11.0cm}}
\vspace{-40mm}
\end{center}
\caption{Bond ($\tau_i$) and torsional ($\alpha_i$) angle distributions
for functional proteins.} 
\label{fig:2}
\end{figure}

In Fig.~\ref{fig:2} the distributions of bond and torsional angles  
are shown together with a scatter plot for these two quantities. Note that the 
calculation of the torsional angle $\alpha_i$ requires four consecutive
\Ca atoms, while the bond angle $\tau_i$ requires only three. The additional
fourth \Ca atom needed for $\alpha_i$ is taken in the N-terminus
direction. 

From Fig.~\ref{fig:2} it is evident that there are strong regularities in
the local structure. The $\tau_i$ and $\alpha_i$ distributions both exhibit a 
clear two-peak structure, which can be associated with two well-populated 
regions in the $(\tau_i,\alpha_i)$-plane. One of these regions, 
$\tau_i \in$ [85$^\circ$,100$^\circ$] and $\alpha_i \in$ [35$^\circ$,70$^\circ$], 
corresponds to right-handed $\alpha$-helix and the other, 
$\tau_i \in$ [105$^\circ$,145$^\circ$] and $\alpha_i \in$ [170$^\circ$,250$^\circ$], 
corresponds to $\beta$-sheet. 

\begin{figure}[htbp]
\begin{center}
\vspace{-35mm}
\mbox{\hspace{-31mm}\psfig{figure=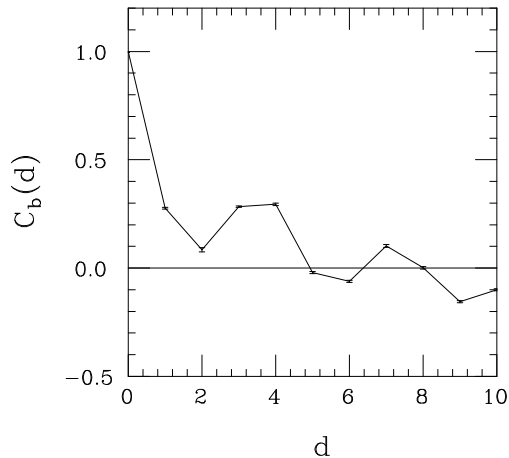,width=10.5cm,height=14cm}
}
\vspace{-50mm}
\end{center}
\caption{The correlation function $C_b(d)$ for 
functional proteins (see Eq.~\protect\ref{corr}).}
\label{fig:3}
\end{figure}

{\bf Local correlations.} We study local correlations using the function 
\beq
C_b(d)={1\over N-d-1}\sum_{i=1}^{N-d-1} \bv_i\cdot\bv_{i+d}
\label{corr}
\eeq  
where $\bv_i$ are normalized (virtual) bond vectors. Local correlations 
along protein chains have been studied by Ref.~\cite{Socci:94a}, using a correlation 
function slightly different from that in Eq.~\ref{corr}. In Fig.~\ref{fig:3}a we 
show the correlation function $C_b(d)$ for functional proteins. As can be seen
from this figure, there are significant correlations at least out to separations of  
about eight residues. The oscillations present can be related to the presence of 
$\alpha$-helix structure, which has a period of 3.6.      

\section{Methods}

\subsection{Monte Carlo Method}

Numerical simulations have been performed for a variety of different choices 
of ($\kappa_1$,  $\kappa_2$) and different sequences and temperatures.
At low temperatures conventional Monte Carlo methods tend to become extremely 
time-consuming, due to the presence of high free-energy barriers. 
As in Refs.~\cite{Irback:95b,Irback:96b} we have therefore chosen to employ 
the dynamical-parameter method, which means that some parameter of the 
model is taken as a dynamical variable which takes values
ranging over a definite set. In this way it is possible to greatly improve 
the frequency of transitions between the different
free-energy valleys; for the 2D model studied in 
Ref.~\cite{Irback:95b} speedup factors of $10^3$--$10^4$
were observed.    

In the present work the temperature is treated as a dynamical variable  
(``simulated tempering'' \cite{Marinari:92}). More precisely, the joint 
probability distribution 
\beq
\label{Pdis}
P(\bv,k)\propto\exp(-g_k-E(\bv,\sigma)/T_k) \,
\eeq
is simulated, where $T_k$, $k=1,\ldots,K$, are the allowed values of the 
temperature.  The $g_k$'s are tunable parameters which must be chosen 
carefully for each sequence. We refer the reader to  
Refs.~\cite{Irback:95b,Irback:96b} for details on how to determine these 
parameters.  The joint distribution $P(\bv,k)$ is simulated by using  
separate Metropolis steps~\cite{Metropolis:53} in $k$ and $\bv$. 
For $\bv$ we use two types of elementary moves: rotations of single bonds    
and moves of pivot type~\cite{Lal:69}.

In our simulations we used a set of $K=25$ allowed temperatures, which
are equidistant in $1/T$ and ranging from 0.15 to 3.0. 

In order to study the energy level spectrum we use a quenching procedure; 
in the course of the simulations the system is quenched to zero 
temperature at regular intervals. For this purpose we employ a conjugate 
gradient method. We found this method more efficient than using a 
Monte Carlo algorithm with $T=0$. Also, we tested two different 
conjugate gradient algorithms. In the conjugate gradient method 
successive minimizations are carried out along different lines.
Information about the gradient is needed to define the lines, 
and may or may not be utilized for the minimizations. For the present 
problem we found that minimization without gradient calculation 
is faster.  The quenching procedure only accounts for a small portion of 
the total computing time.

\begin{figure}[tbp]
\begin{center}
\vspace{-45mm}
\mbox{\hspace{0mm}\psfig{figure=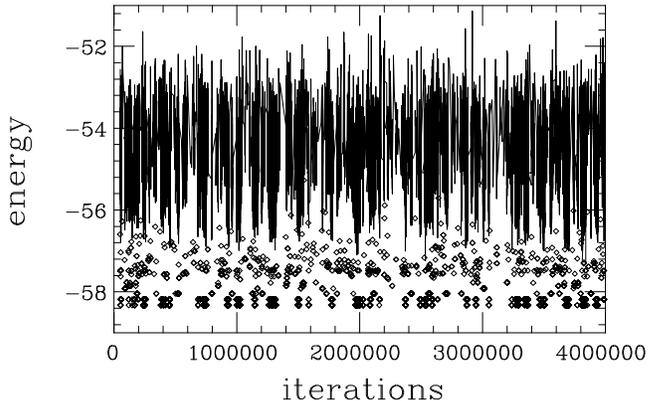,width=10.5cm,height=14cm}}
\vspace{-45mm}
\end{center}
\caption{Evolution of the quenched (diamonds) and unquenched (line) energies
in the simulation of sequence 1 (see Table 1). Measurements were
taken every 10 iterations. Shown are the data corresponding to the lowest allowed
temperature. The thermalization phase
 of 50000 sweeps is not shown.}
\label{fig:4}
\end{figure}

In Fig.~\ref{fig:4} we show the evolution of the quenched and unquenched energies
in a typical simulation of a $N=20$ chain.

\subsection{Measurements}

In Sec.~3 we discussed measurements of local properties of the chains.
In order to study the stability of the full native structure, further 
information is needed. To this end we have studied the distribution of
the mean-square distance between configurations, $\delta_{ab}^2$. For
two configurations $a$ and $b$, $\delta_{ab}^2$ is defined as  
\beq
\delta^2_{ab} = \min {1\over N}\sum_{i=1}^N 
|\bar x^{(a)}_i-\bar x^{(b)}_i|^2
\label{delta}
\eeq
where $|\bar x^{(a)}_i-\bar x^{(b)}_i|$ denotes the distance 
between the sites $\bar x^{(a)}_i$ and $\bar x^{(b)}_i$, 
and where the minimum is taken over translations, rotations and reflections.
The corresponding distribution, $P(\delta^2)$, for fixed temperature and 
sequence, reads
\beq
P(\delta^2)= \frac{1}{Z(T;\sigma)^{2}}\int 
d\bv^{(a)}d\bv^{(b)}
\delta(\delta^2-\delta_{ab}^2)
{\rm e}^{-E(\bv^{(a)};\sigma)/T)}
{\rm e}^{-E(\bv^{(b)};\sigma)/T)} 
\label{Pd}
\eeq
where $\delta(\cdot)$ denotes the Dirac delta function. For convenience,
we often use the mean
\beq
\langle \delta^2 \rangle = \int d\delta'^2 P(\delta'^2)\delta'^2
\eeq
rather than the full distribution $P(\delta^2)$.  

We have also measured the specific heat $C_V$ and gyration
radius $\rgyr$, defined by    
\bea
C_V&=&{1\over T^2}\big(\ev{E^2}-\ev{E}^2\big)\\
\rgyr^2&=&{1\over N}\sum_{i=1}^N \big(\ev{\bar x_i^2}-\ev{\bar x_i}^2\big) 
\label{C_rgyr}
\eea
The specific heat has a maximum in the vicinity of the folding 
transition. To accurately determine the height and location of 
this maximum we use the multihistogram method~\cite{Swendsen:88,Swendsen:89}.

In addition to these measurements, we employ the quenching procedure 
described above. Removing the thermal noise in this way is of great help in 
monitoring the evolution of the system, but requires a substantial amount of 
computer time. Therefore, we have performed these calculations at larger intervals 
than other measurements, and only at the lowest of the temperatures studied.

The quenching procedure provides us with a set of low-lying local 
energy minima. For some of the studied chains we believe that the lowest 
minimum obtained in our simulation is the global minimum of the 
energy function, as will be discussed below. 
The mean-square distance to the global minimum will be 
denoted by $\delta_0^2$ (cf Eq.~\ref{delta}). Using the 
corresponding distribution, $P(\delta_0^2)$, we define a probability $p_0$ 
of finding the system within a small neighborhood of the global minimum. We take
\beq
p_0=\int_0^\Delta P(\delta_0^2)d\delta_0^2
\label{p0}
\eeq   
with the parameter $\Delta=0.04$; this choice of $\Delta$ is motivated by the
$\delta^2_0$ distributions (cf Figs.~\ref{fig:11} a and b). Let us stress that
the distribution $P(\delta_0^2)$ is different from the distribution 
$P(\delta^2)$ introduced earlier; $P(\delta^2)$ measures general 
fluctuations rather than deviations from a given state.

The degree of folding may be defined in terms of the quantity $p_0$.  
Alternatively, it may be defined as 
\beq
Q={n^o\over n} 
\label{Q}
\eeq
where $n^o$ is the number of occupied native contacts, and $n$ is the total
number of native contacts. Two monomers $i$ and $j$ are taken to be in
contact if $r_{ij}^2<1.75$; this cutoff is motivated by the distribution
$P(r_{ij}^2)$ (not displayed in this paper). We have
$Q=1$ for the native structure. Also, it is
useful to divide the set of native contacts into local and global contacts.        
A contact between monomers $i$ and $j$ will be called local if 
$2\le |i-j|\le 4$ and global if $|i-j|>4$. We set $Q_l=n^o_l/n_l$, where
$n^o_l$ is the number of occupied local native contacts and $n_l$ is the total
number of local native contacts. Similarly, we define $Q_g$ as the
fraction of occupied global native contacts.
 
\section{Results}

The model is defined by two parameters, $\kappa_1$ and $\kappa_2$, 
which set the strengths of the local interactions. In order to investigate the
importance of these interactions, we first performed preliminary
runs for a number of different $(\kappa_1,\kappa_2)$ values. In particular, 
these explorations aimed at establishing the balance between the local and 
global interactions --- the overall scale of  $(\kappa_1,\kappa_2)$.
More extensive simulations were then carried out for three different choices  
\beq
 (\kappa_1,\kappa_2)=
                             \left\{ \begin{array}{ll}
                                & \mbox{(0,0)}\\
                                &\\
                                & \mbox{(-1,0)}\\
                                &\\
                                & \mbox{(-1,0.5)}
                              \end{array} \right.\\
\eeq
using sequences of length $N=20$. It turns out that shorter chains exhibit 
less interesting and discriminative behavior.  The sequences, which were 
deliberately chosen to represent a variety of behavior, are listed in 
Table \ref{tab:1}.
\begin{table}[tb]
\begin{center}
\begin{tabular}{|c|c|c c c|} \hline
no.      & sequence                       &$T_s$ & $C_{V}^{\max}$ & $T_f$ \\
\hline
1        & $BAAA\ AAAB\ AAAA\ BAAB\ AABB$ & 0.361 &   51.4    & $<0.15$ \\
2        & $BAAB\ AAAA\ BABA\ ABAA\ AAAB$ & 0.319 &   55.9    & $<0.15$ \\
3        & $AAAA\ BBAA\ AABA\ ABAA\ ABBA$ & 0.298 &   65.8    & 0.23    \\
4        & $AAAA\ BAAB\ ABAA\ BBAA\ ABAA$ & 0.273 &   61.5    & 0.22    \\
5        & $BAAB\ BAAA\ BBBA\ BABA\ ABAB$ & 0.327 &   49.8    & $<0.15$ \\
6        & $AAAB\ BABB\ ABAB\ BABA\ BABA$ & 0.257 &   62.6    & 0.15    \\
\hline
\end{tabular}
\caption{The six sequences studied. The errors in $T_s$ and $C_{V}^{\max}$
are less than $0.005$ and $0.3$ respectively. The errors in $T_f$
are approximately $0.02$.  }
\label{tab:1}
\end{center}
\end{table}

In this section we first compare the behavior of the model at the 
three values of $(\kappa_1,\kappa_2)$ with respect to the local interactions 
and low $T$ properties in order to single out one set of  
$(\kappa_1,\kappa_2)$ values. We then examine the folding properties 
and thermodynamic behavior for  $(\kappa_1,\kappa_2)=(-1,0.5)$ in some detail.

\subsection{The Local Interactions}

\begin{figure}[tb]
\vspace{0.7in}
\begin{center}
\vspace{-55mm}
\mbox{\hspace{0mm}\psfig{figure=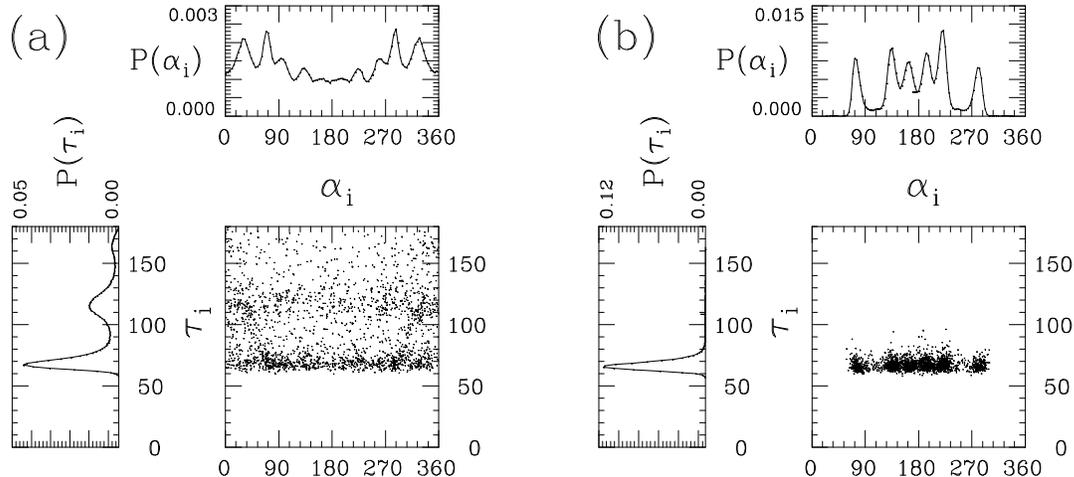,width=10.5cm,height=14.5cm}}
\vspace{-50mm}
\end{center}
\caption{Bond ($\tau_i$) and torsional ($\alpha_i$) angle distributions at $T=0.15$ for 
the six simulated sequences, {\bf (a)} pure Lennard-Jones potential 
$(\kappa_1,\kappa_2)=(0,0)$ and {\bf (b)}  $(\kappa_1,\kappa_2)=(-1,0.5)$.
The potential corresponding to $(\kappa_1,\kappa_2)=(-1,0)$ yields a similar 
distribution as in (b).} 
\label{fig:5}
\end{figure}

\begin{figure}[bt]
\begin{center}
\vspace{-35mm}
\mbox{\hspace{-31mm}\psfig{figure=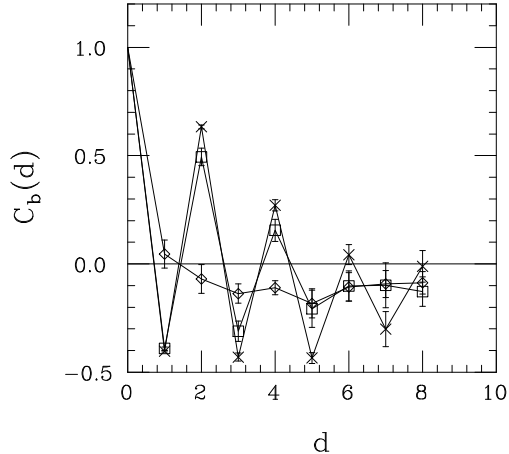,width=10.5cm,height=14cm}}
\vspace{-50mm}
\end{center}
\caption{The correlation function $C_b(d)$ (Eq.~\protect\ref{corr}) for 
3D model chains; $(\kappa_1,\kappa_2)$=(0,0) $(\diamond)$, (-1,0) $(\Box)$ 
and (-1,0.5) $(\times)$.} 
\label{fig:6}
\end{figure}

Prior to investigating folding properties for different choices of $(\kappa_1,\kappa_2)$, 
we compare the bond and torsional angle distributions and the local correlations of 
the chains with those of functional proteins. 
In Figs.~\ref{fig:5} and \ref{fig:6} the model counterparts of 
Figs.~\ref{fig:2} and \ref{fig:3} are shown. The data in Fig.~\ref{fig:5}
were obtained at a fixed temperature, $T=0.15$,  
while those in Fig.~\ref{fig:6} were obtained using quenched 
low energy structures. In both cases we expect the data to reflect the  
behavior for a wide range of not too high temperatures.

The results strongly indicate the 
need for local interactions when it comes to mimicking functional proteins; 
the strong regularities in the local structure observed for proteins 
are to a large extent missing for a pure Lennard-Jones potential. 
This can be seen from the torsional angle distribution and the correlation 
function $C_b(d)$. The torsional angle $\alpha_i$ varies over the whole range of 
$360^\circ$, without any strongly suppressed regions, and the correlation $C_b(d)$ is very 
weak for all $d>0$. The conclusion that local correlations 
are weak for a pure Lennard-Jones potential is in good agreement with the findings 
of Ref.~\cite{Socci:94a}.   


From the Figs.~\ref{fig:5} and \ref{fig:6} it is also clear that changing 
$\kappa_1$ from 0 to -1 leads to a significant improvement; the torsional angle 
distribution becomes concentrated to a few relatively small regions, 
and local correlations become stronger. The range of the correlation 
$C_b(d)$ is for $\kappa_1=-1$ comparable to what it is for proteins. Needless 
to say, the model is not intended to reproduce the precise form of the correlations.  
However, it is encouraging that the qualitative behavior of this very simple 
model is consistent with the one from functional proteins. 

The remaining question is whether a non-zero $\kappa_2$ is called for. 
No conclusive evidence can be drawn from Figs.~\ref{fig:5} and 
\ref{fig:6} alone in this respect, although one may argue that 
the range of the correlation $C_b(d)$ is still somewhat short for 
$(\kappa_1,\kappa_2)=(-1,0)$. 

\begin{figure}[tbp]
\begin{center}
\vspace{-35mm}
\mbox{\hspace{-30mm}\psfig{figure=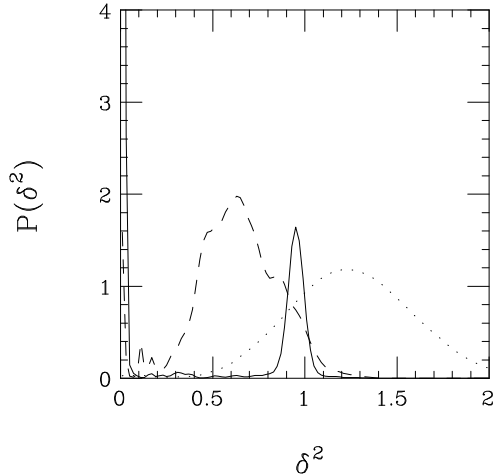,width=10.5cm,height=14cm}}
\vspace{-50mm}
\end{center}
\caption{$P(\delta ^2)$ at $T=0.15$ for sequence 4 (cf Table 1);
($\kappa_1$,$\kappa_2$)=(0,0) (dots), (-1,0) (dashes), and  
(-1,0.5) (solid). The low-$\delta^2$ peak for ($\kappa_1$,$\kappa_2$)=(-1,0.5) extends to 37.} 
\label{fig:7}
\end{figure}

Next we investigate how the choice of parameters affects the key folding property 
identified in Ref.~\cite{Irback:96b} --- the $\delta^2$ distribution should be 
peaked for low $\delta^2$. In Fig.~\ref{fig:7},  $P(\delta^2)$ is shown for sequence 
4 in Table~\ref{tab:1}. In contrast to the local properties discussed above, 
$P(\delta^2)$ is strongly sequence dependent. Nevertheless, Fig.~\ref{fig:7} 
illustrates some general trends seen in the data. First, the pure 
Lennard-Jones interaction yields a very broad distribution of $\delta^2$, implying
that folding properties are poor. This is true for all the six sequences 
studied. Second, although the behavior is different for one of the six sequences, 
the $\kappa_2\ne 0$ choice appears to have a distinct edge over the one 
ignoring the torsional angle interaction.

In summary, our results show that local interactions are essential in order
to get a regular local structure and structural stability. When comparing
the results for $(\kappa_1,\kappa_2)=(-1,0)$ and (-1,0.5), we find that
the structural stability tends to be highest for $(\kappa_1,\kappa_2)=(-1,0.5)$. 
In what follows we will focus on this choice of parameters for the six sequences in 
Table~\ref{tab:1}.

\subsection{2D Revisited}

In Refs.~\cite{Irback:95b,Irback:96b}, a similar model \cite{Stillinger:93} 
was studied in 2D using somewhat different Lennard-Jones parameters 
and local interactions corresponding to 
($\kappa_1,\kappa_2$)=($\frac{1}{4},0$) (cf Eq.~\ref{energy}). 
Here, we briefly discuss the importance of local interactions for this 
2D model.

For this purpose, we leave out the local interaction term, i.e. we set
($\kappa_1,\kappa_2$)=($0,0$). Using this pure Lennard-Jones potential, we
redo the simulations for 15 sequences. 
At $T=0.15$, we find a strong correlation between the $\ev{\delta^2}$'s for the
two types of potentials; thus $\ev{\delta^2}$ varies widely with sequence   
even for the ($\kappa_1,\kappa_2$)=($0,0$) potential. This is in contrast to 
our findings in 3D, where all the sequences studied have a large 
$\ev{\delta^2}$ for ($\kappa_1,\kappa_2$)=($0,0$).

\subsection{Folding Properties}  

When investigating the folding properties of the chains we focus on 
the thermodynamics. This is in part inspired by the results from 
Ref.~\cite{Irback:96b},  where the thermodynamic properties exhibited 
strong sequence dependence. Initially we examine the various thermodynamic 
quantities defined in Sec. 4.2 over the entire probed $T$ range. Then 
we proceed with the ``magnifying glass'' to the 0.15$<T<$0.50 region, 
where the different chains exhibit strongest difference in behavior and  
the folding properties can be studied in some detail.

\subsubsection{Thermodynamic Behavior}

\begin{figure}[ht]
\begin{center}
\vspace{-52mm}
\mbox{\hspace{-31mm}\psfig{figure=fig8a.ps,width=10.5cm,height=14cm}
\hspace{-30mm}      \psfig{figure=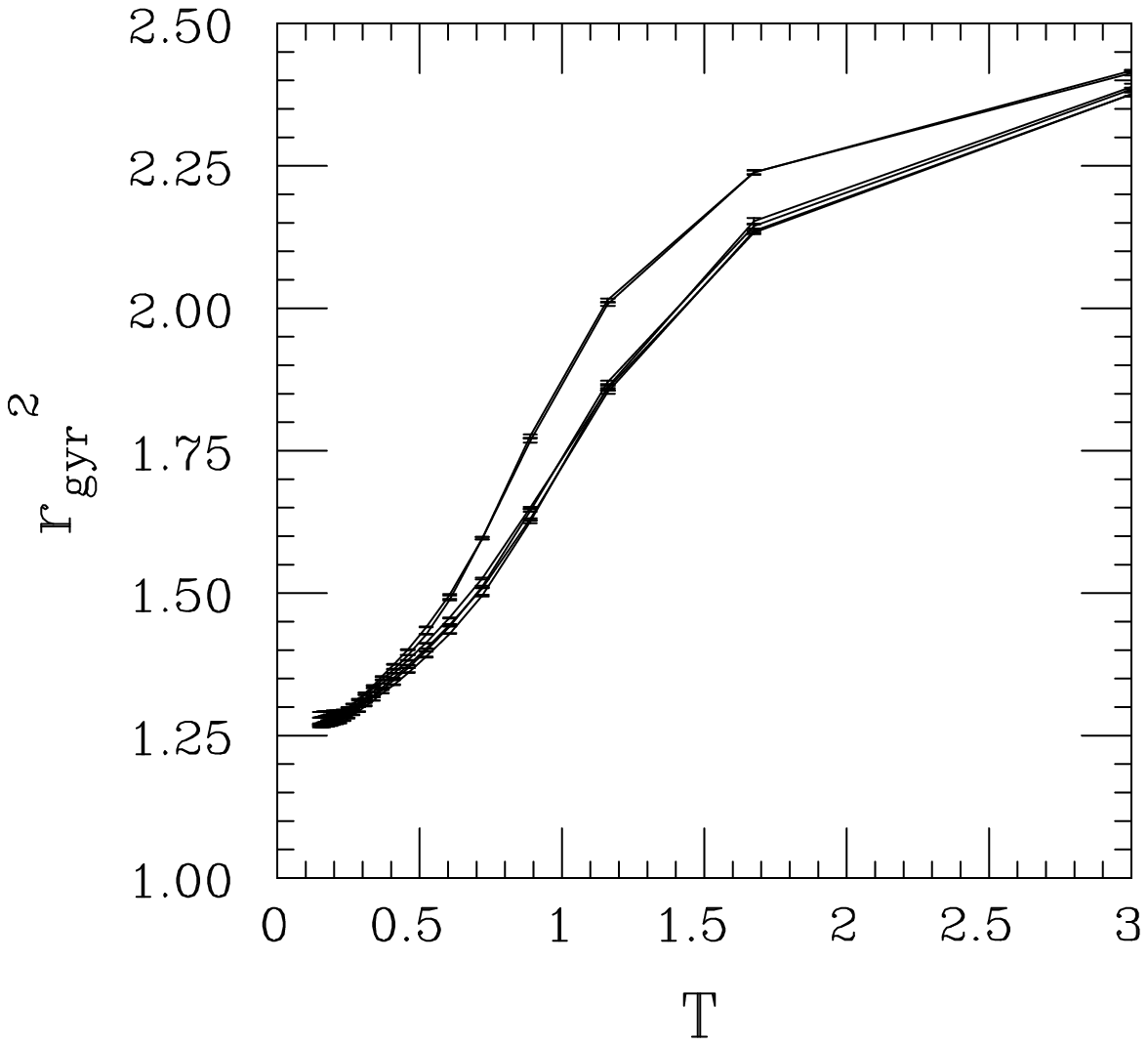,width=10.5cm,height=14cm}}
\vspace{-45mm}
\vspace{-35mm}
\mbox{\hspace{-31mm}\psfig{figure=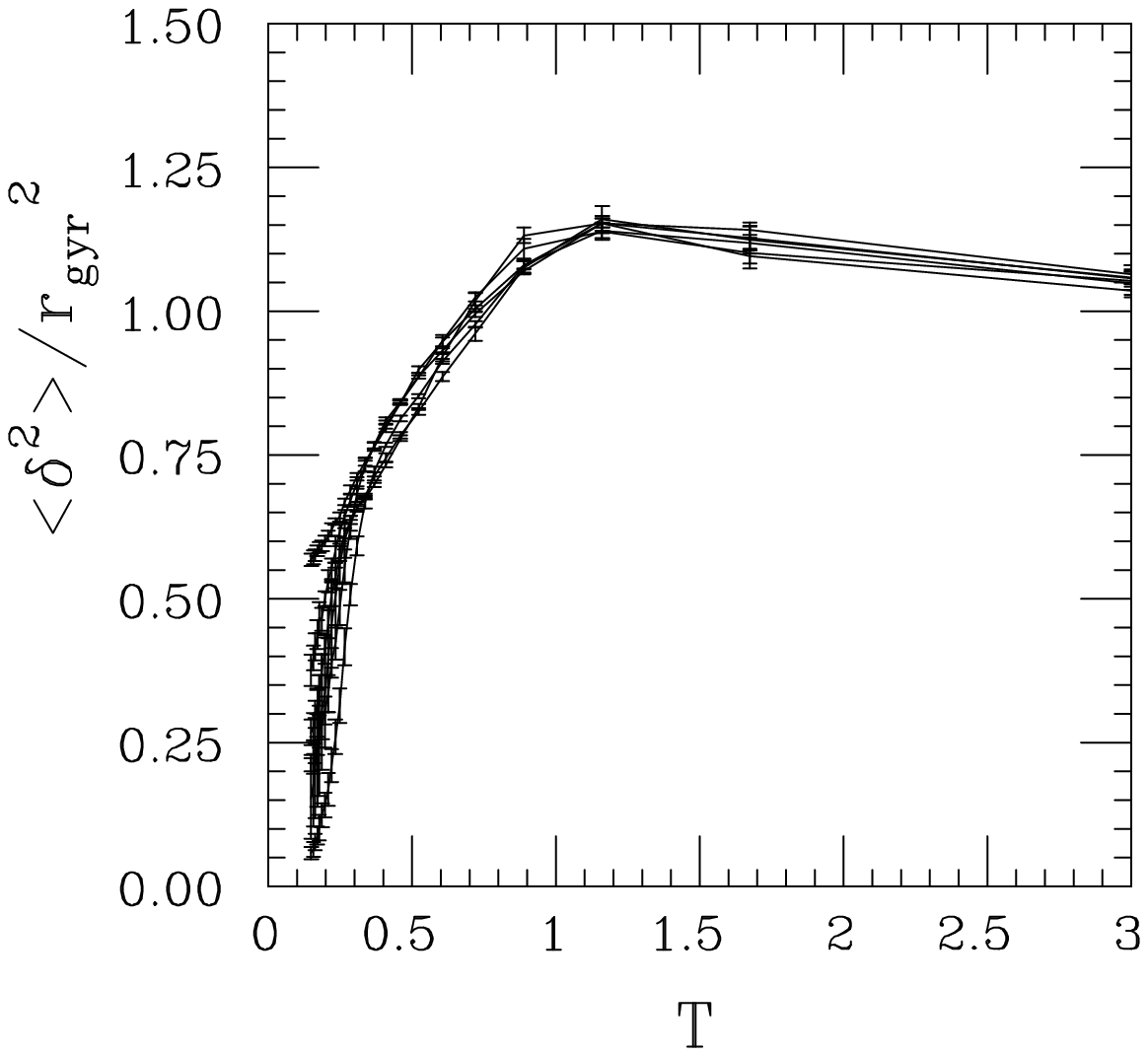,width=10.5cm,height=14cm}
\hspace{-30mm}      \psfig{figure=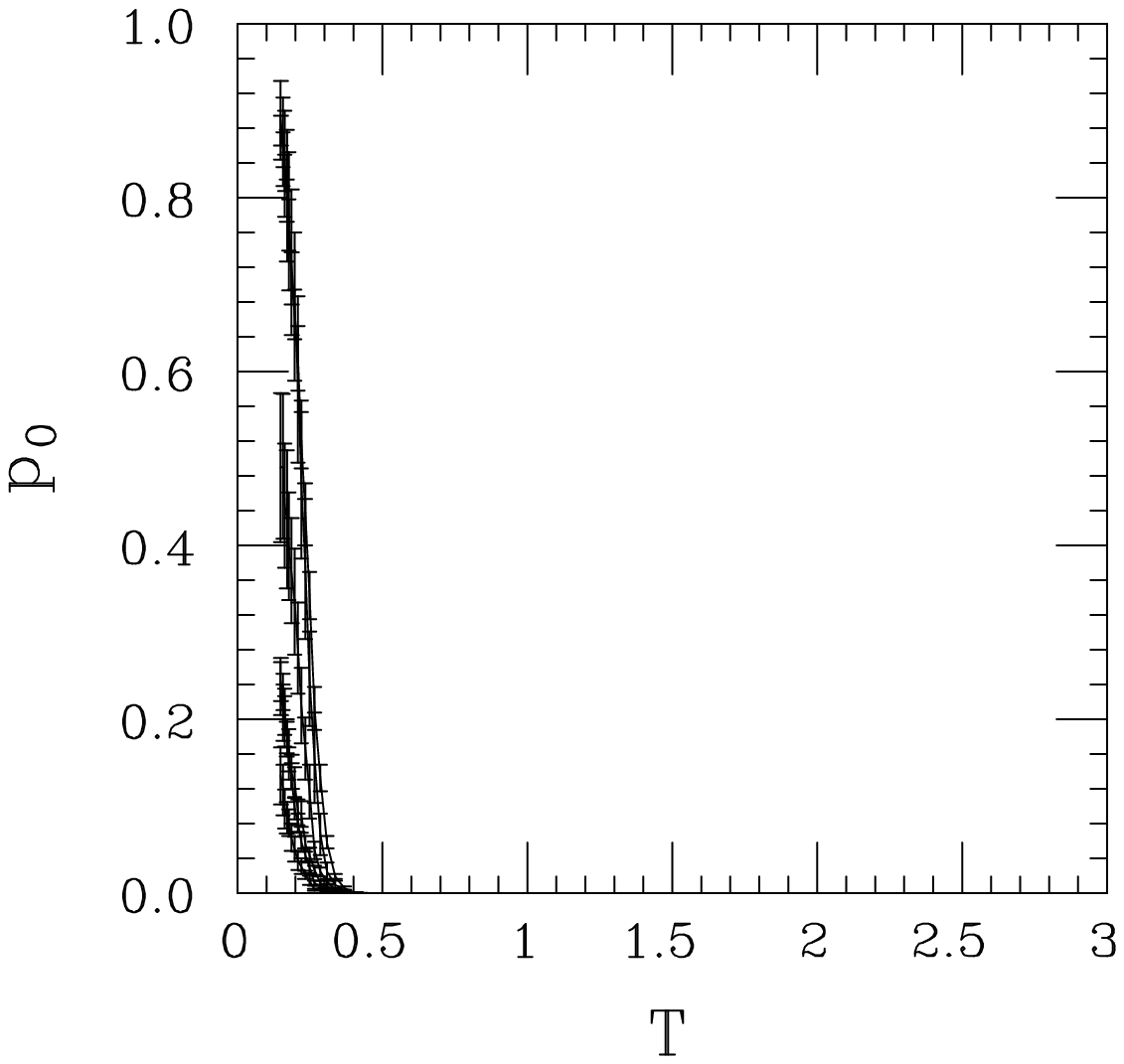,width=10.5cm,height=14cm}}
\vspace{-45mm}
\vspace{-35mm}
\mbox{\hspace{-31mm}\psfig{figure=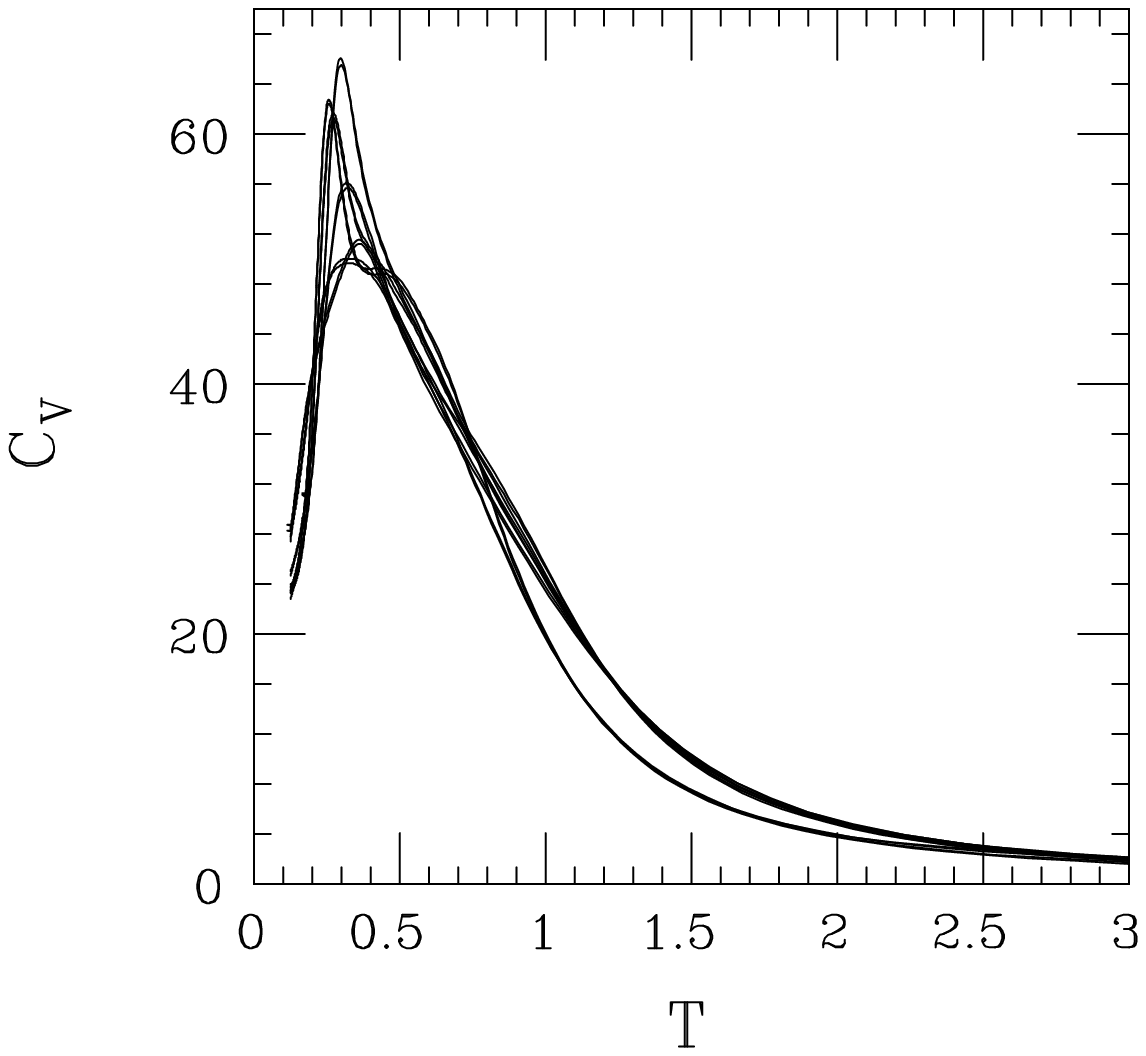,width=10.5cm,height=14cm}
\hspace{-30mm}      \psfig{figure=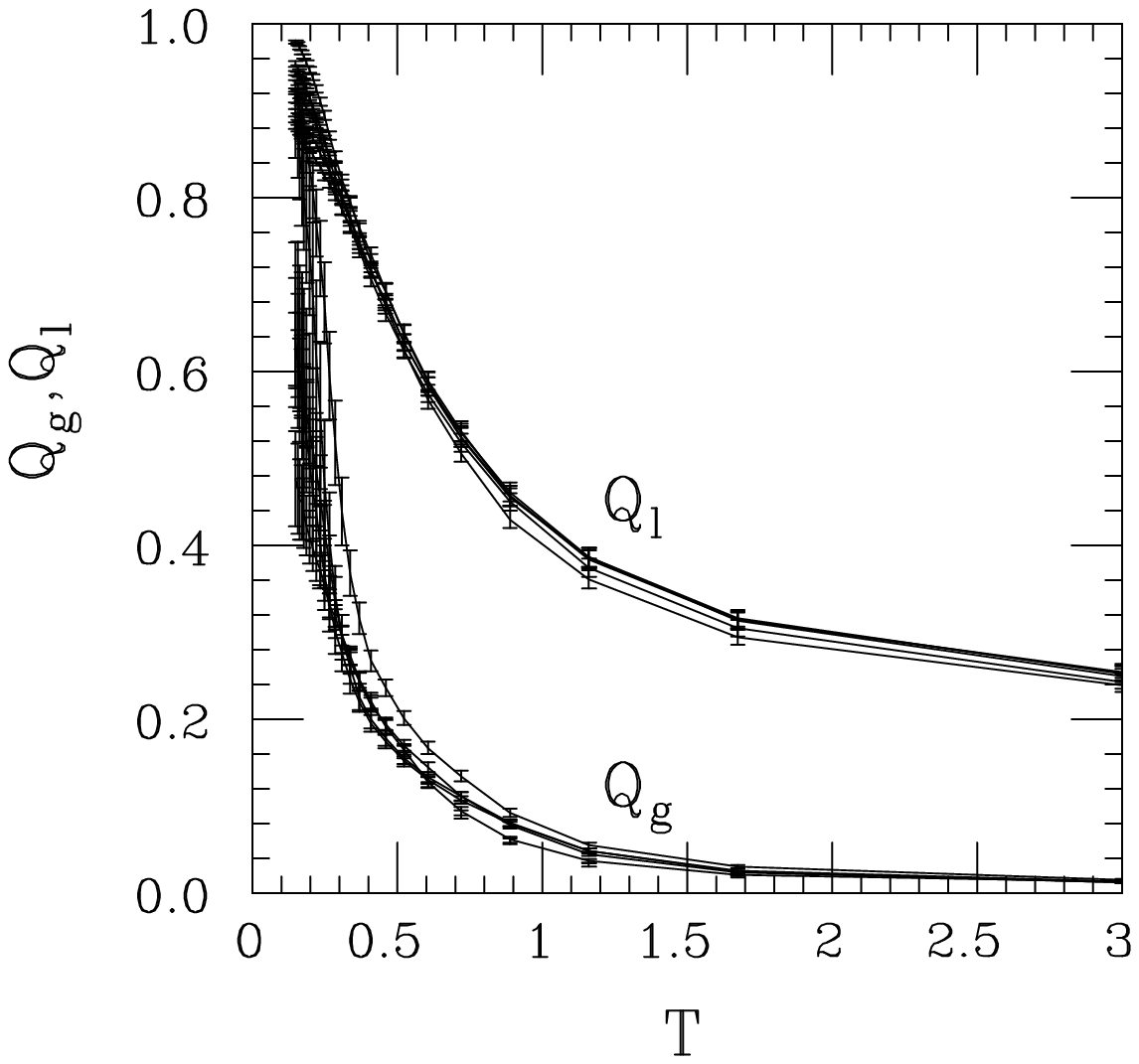,width=10.5cm,height=14cm}}
\vspace{-45mm}
\end{center}
\caption{ Thermodynamic properties for the six chains in Table 1 using  ($\kappa_1,\kappa_2$)=(-1,0.5).}
\label{fig:8}
\end{figure}

In Fig.~\ref{fig:8} the behavior of the different thermodynamic 
quantities are shown over the entire $T$ range. The overall size of the 
molecule, as measured by $\rgyr^2$, decreases substantially when $T$ decreases 
from 3 to 0.15, as can be seen from Fig.~\ref{fig:8}. This decrease 
is gradual. The data points essentially fall onto two different curves. 
The upper curve corresponds to the sequences 5 and 6 with composition 
10A+10B, and the lower curve to the sequences  1--4 with composition  
14A+6B.
 
\begin{figure}[t]
\begin{center}
\vspace{-82mm}
\mbox{
\hspace{-31mm}
\psfig{figure=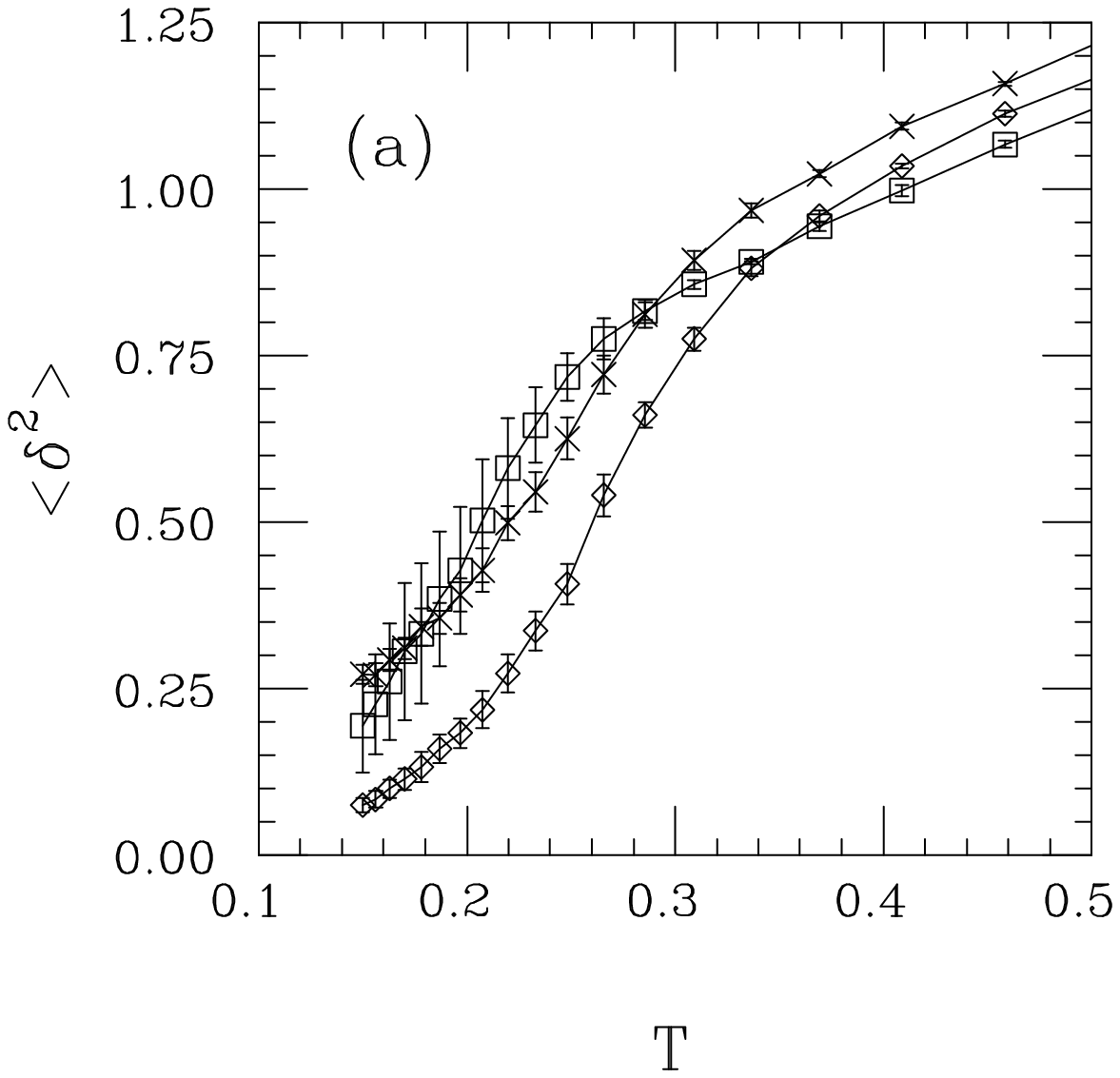,width=10.5cm,height=14cm}
\hspace{-31mm}      
\psfig{figure=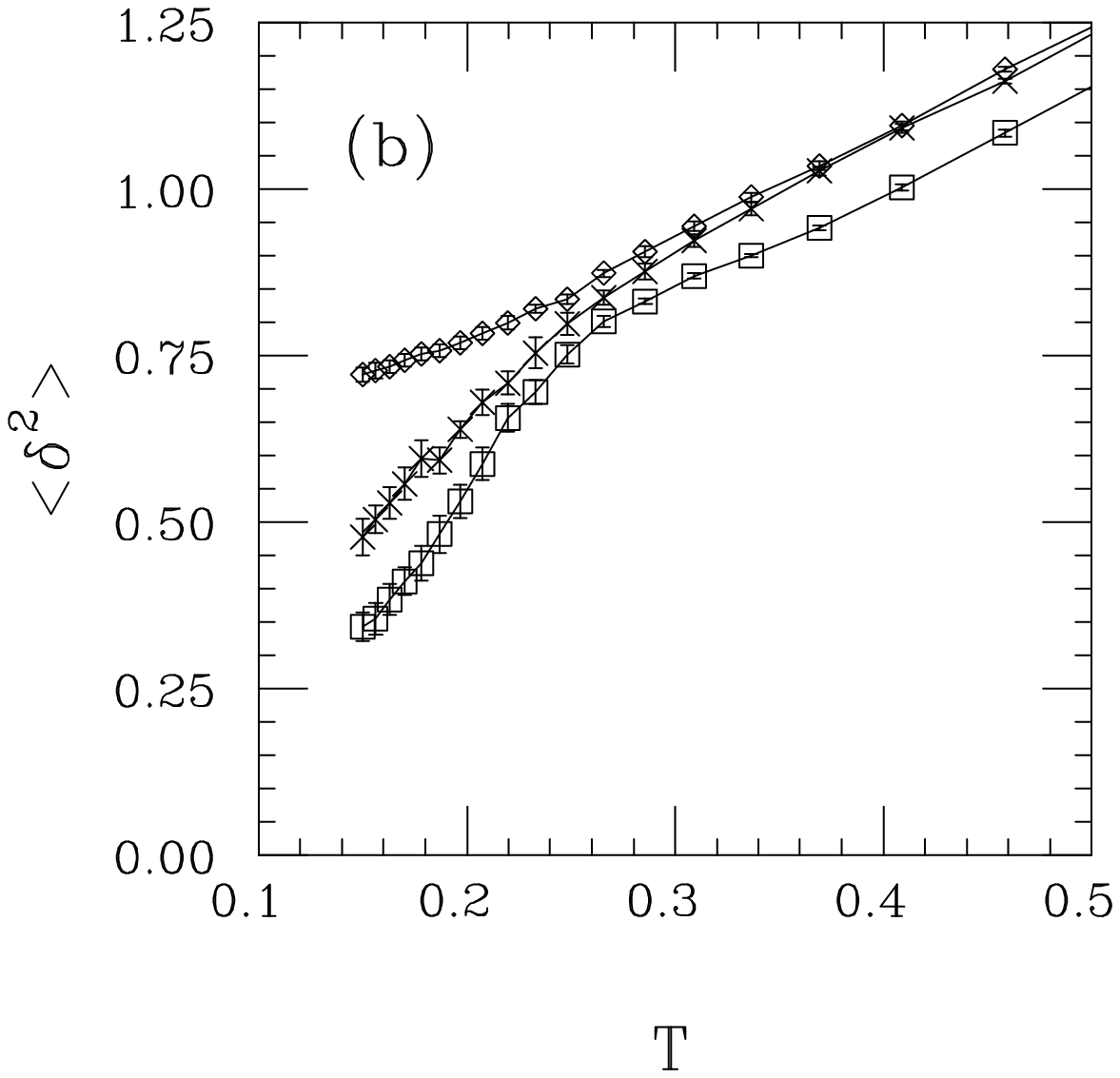,width=10.5cm,height=14cm}}
\vspace{-5mm}
\end{center}
\caption{$\ev{\delta^2}$ as a function of $T$ 
for the sequences of Table 1: {\bf (a)} 4 ($\Box$), 2  ($\times $)  
and 3 ($\diamond $). {\bf (b)} 6 ($\Box$), 1 ($\times $) and 
5 ($\diamond $).}
\label{fig:9}
\end{figure}

Next we turn to $\ev{\delta^2}$, which measures the size of the fluctuations. 
In Fig.~\ref{fig:8} we show the relative magnitude 
$\ev{\delta^2}/\rgyr^2$. This ratio 
exhibits a peak slightly above $T=1$, and is approximately sequence 
independent down to $T\approx 0.4$; above this temperature $\ev{\delta^2}$ 
shows a sequence (composition) dependence similar to that of $\rgyr^2$.
Below $T\approx 0.4$ the situation is different.
Here $\ev{\delta^2}$ is strongly sequence dependent, in contrast to $\rgyr^2$. 
In Fig.~\ref{fig:9} the results are shown for $\ev{\delta^2}$  in the 
low-$T$ region. 

\clearpage
    
\begin{figure}[t]
\begin{center}
\vspace{-30mm}
\mbox{
\hspace{-31mm}
\psfig{figure=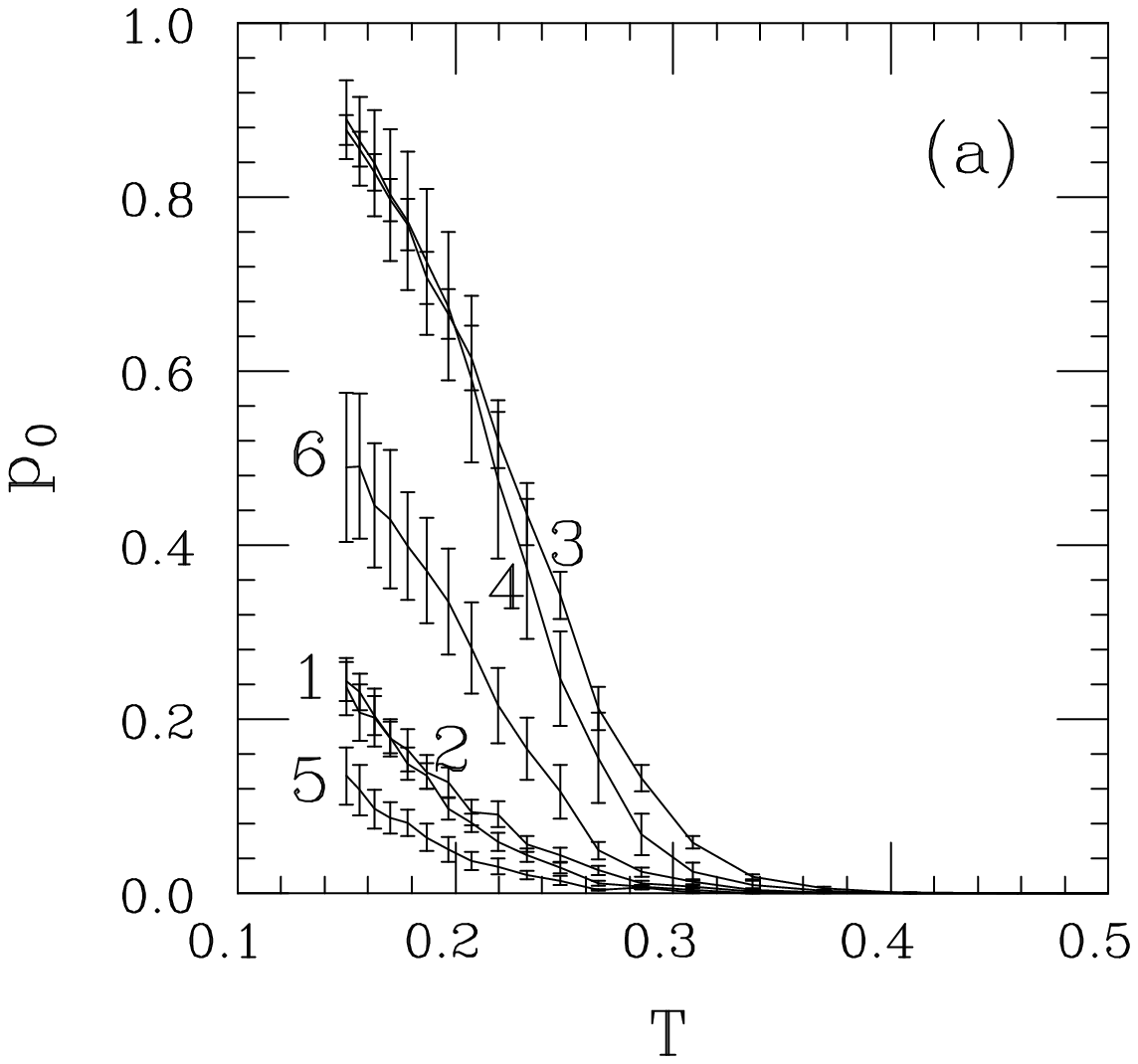,width=10.5cm,height=14cm}
\hspace{-31mm}
\psfig{figure=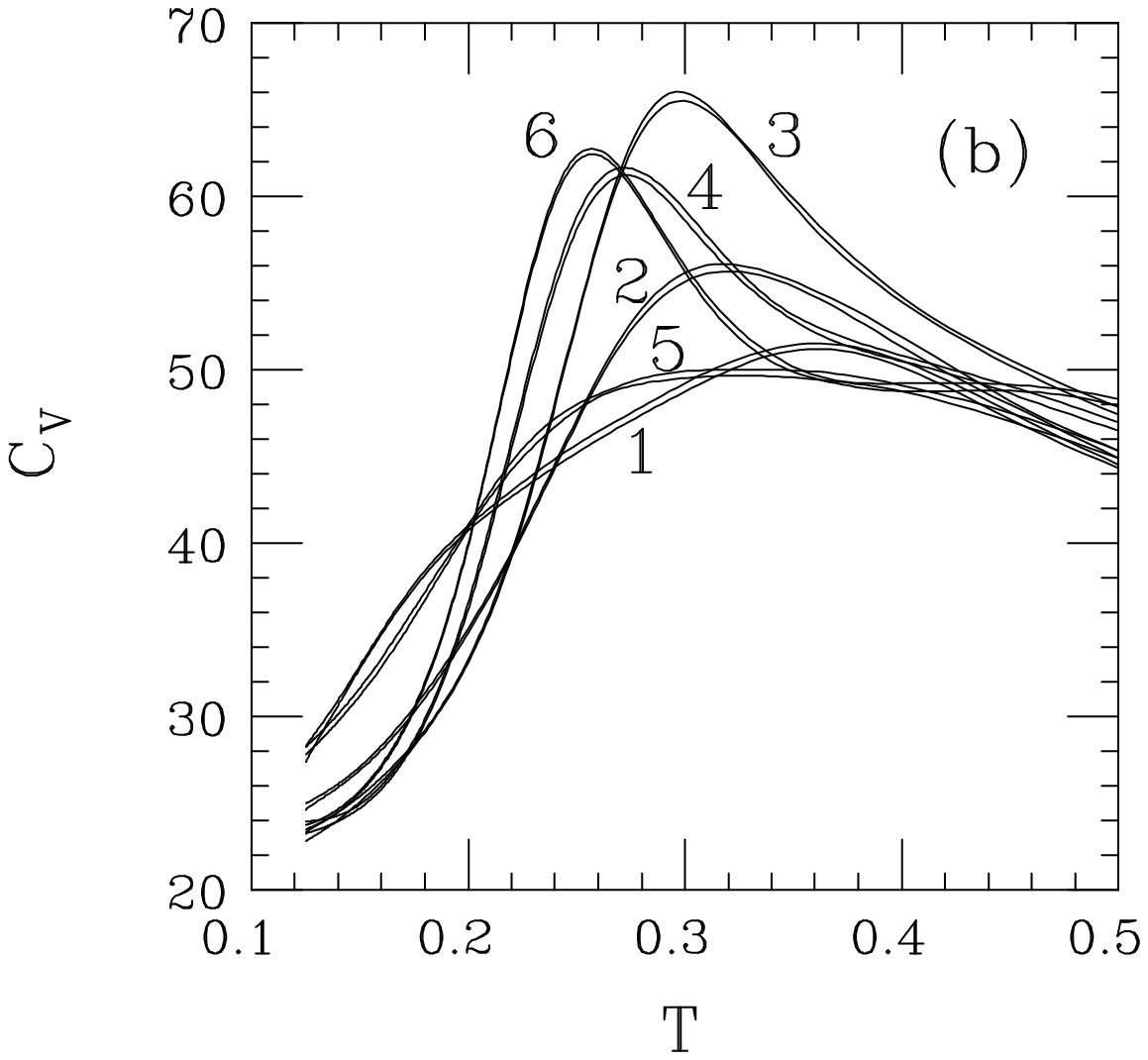,width=10.5cm,height=14cm}}
\vspace{-55mm}
\end{center}
\caption{Enlargements of {\bf (a)} $p_0$ and {\bf (b)} $C_V$ in 
Fig.~\protect\ref{fig:8}.
The bands in (b) correspond to $C_V\pm\sigma$, 
where $\sigma$ is the standard deviation.} 
\label{fig:10}
\end{figure}

In order to understand the low-$T$ behavior it is useful to consider
$p_0$, which measures the relative population of the lowest energy state
(Eq.~\ref{p0}). A comparison of the data for $\ev{\delta^2}$ and $p_0$ shows 
that small $\ev{\delta^2}$ values are associated with large $p_0$ values. 
Therefore, it is reasonable to define the folding temperature $T_f$ as
the temperature where $p_0=1/2$. Estimates of $T_f$ 
are given in Table 1. For three of the six sequences $T_f$ is smaller  
than 0.15, as can be seen from Fig.~\ref{fig:10}a. It should be stressed 
that the shape of the molecule is very compact already above $T_f$. 
This will be discussed in more detail below for one of the sequences. 

From Figs.~\ref{fig:8} and \ref{fig:10}b it can be seen that
the specific heat exhibits a peak in the low-$T$ region,
slightly above $T_f$. The height, $C_{V}^{\max}$, and location, $T_s$, 
of the peak are given in Table 1. We find that $C_{V}^{\max}$ is  
large for the sequences with high $T_f$, which   
is in accordance with the results from the lattice model study of Ref.~\cite{Socci:95}. 
However, our results for $T_s$ are somewhat different from those of  
Ref.~\cite{Socci:95}.  In our model  $T_s$ is
relatively low for sequences with high $T_f$, while the results reported
by Ref.~\cite{Socci:95} show the opposite behavior. Thus, the
separation $T_s-T_f$ exhibits a stronger sequence dependence in our 
model.     

Also shown in Fig.~\ref{fig:8} is the $T$ dependence for the occupancy 
of local and global native contacts. In the vicinity of $T_f$, $Q_l$ is large 
and varies slowly, whereas $Q_g$ changes rapidly. In particular, this demonstrates  
that the formation of local native structure, like the compactification,  
is a gradual process that to a large extent takes place above $T_f$.     

\subsubsection{The Folding Transition Region}  

In this subsection we study the behavior at the  
folding temperature $T_f$, where $p_0=1/2$, in some more detail.
Two different sequences are considered, 4 and 6. 
Sequence 4 has a fairly high $T_f$ value, 
whereas sequence 6 represents a more typical $T_f$. 

\begin{figure}[tp]
\begin{center}
\vspace{-45mm}

\mbox{
\hspace{-30mm}
\psfig{figure=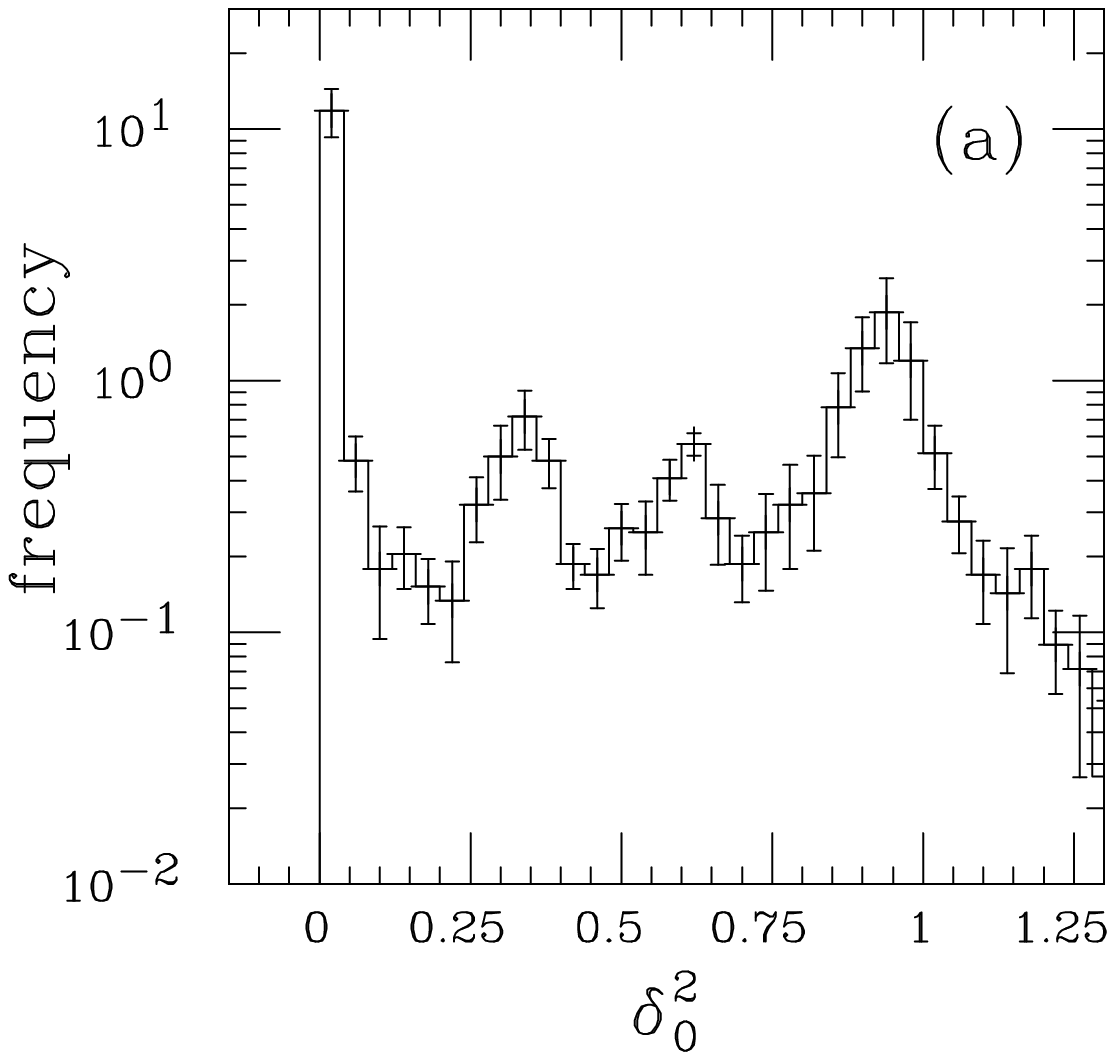,width=10.5cm,height=14cm}
\hspace{-30mm}
\psfig{figure=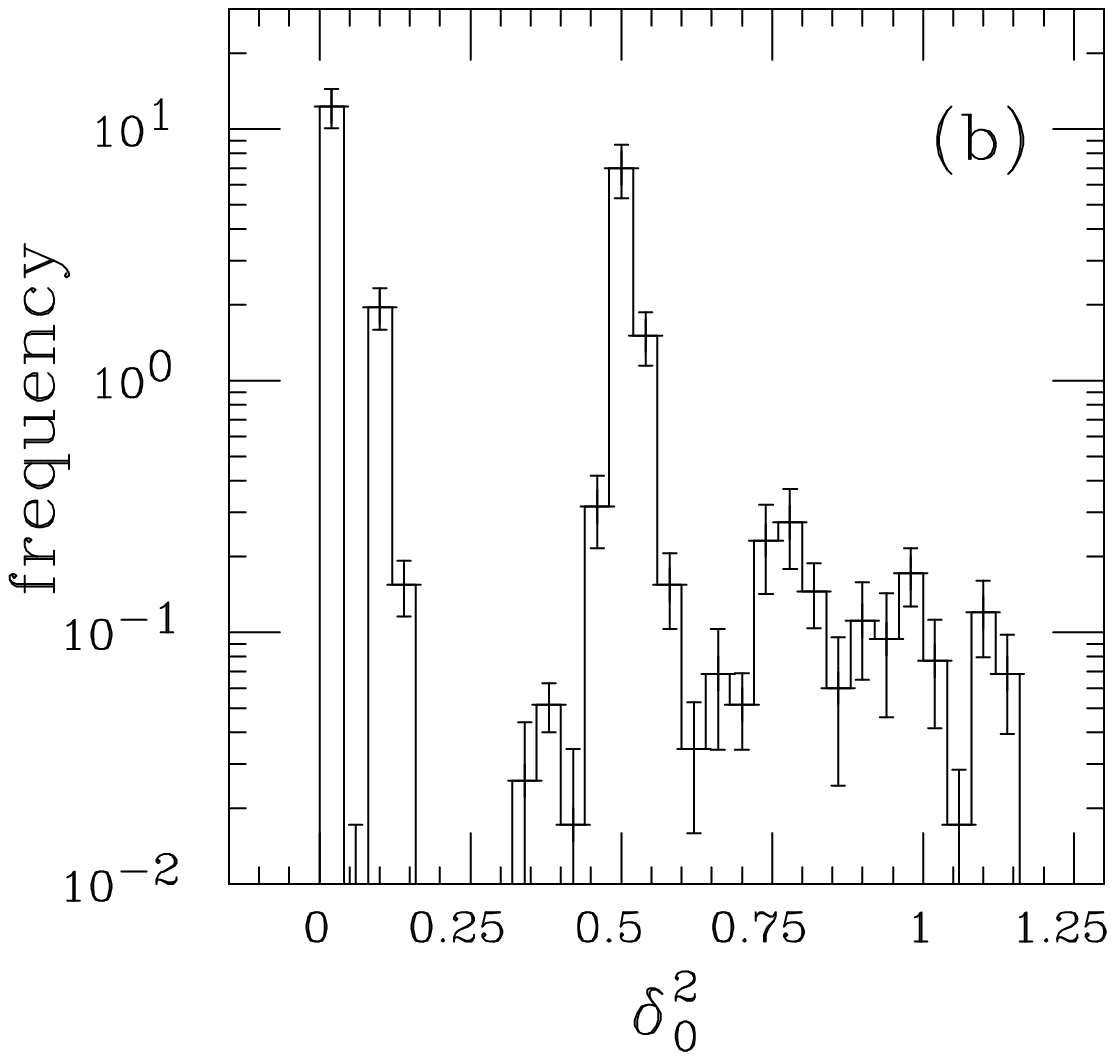,width=10.5cm,height=14cm}
}

\vspace{-70mm}

\mbox{
\hspace{-30mm}
\psfig{figure=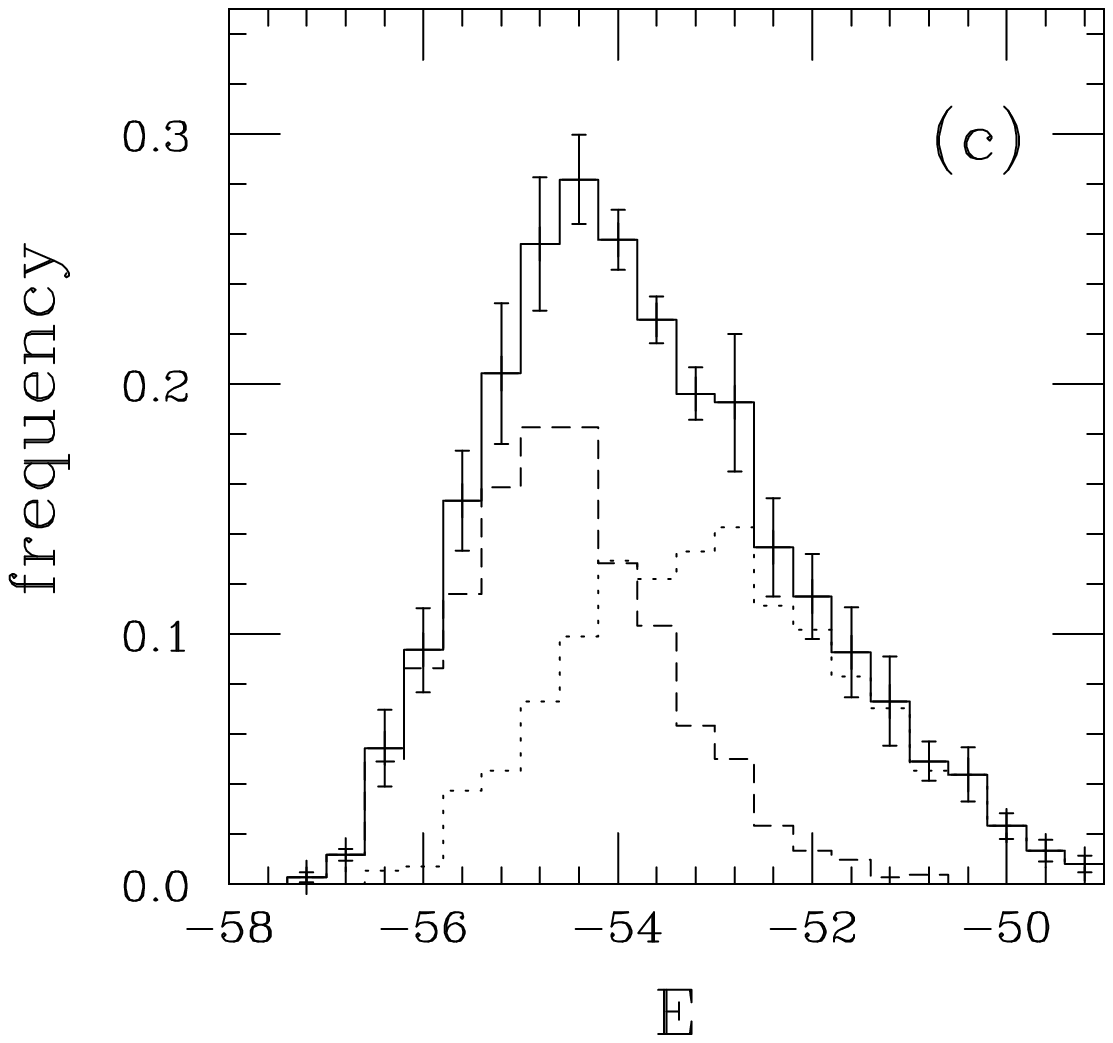,width=10.5cm,height=14cm}
\hspace{-30mm}
\psfig{figure=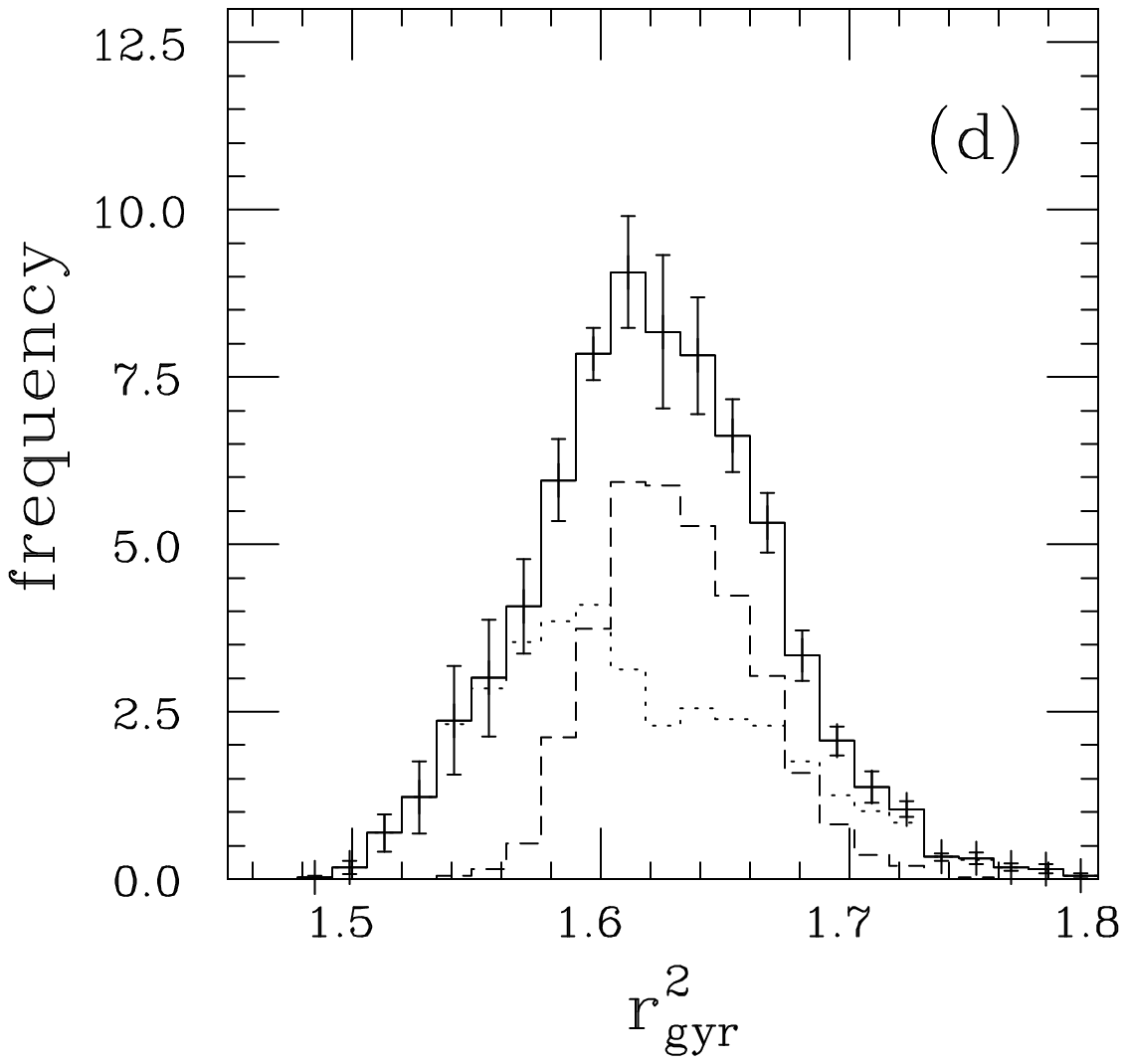,width=10.5cm,height=14cm}
}

\vspace{-45mm}
\end{center}

\caption{Histograms at $T\approx T_f$ for sequence 4 ($T=0.22$) 
and 6 ($T=0.15$).
{\bf (a)} $\delta_0^2$ for seq. 4. 
{\bf (b)} $\delta_0^2$ for seq. 6.
{\bf (c)} $E$ for seq. 4. 
{\bf (d)} $\protect\rgyr^2$ for seq. 4.
The dashed and dotted lines in (c) and  (d) represent the 
contributions corresponding to $\delta_0^2<0.04$ and $\delta_0^2>0.04$,
respectively.} 
\label{fig:11}
\end{figure}

In Figs.~\ref{fig:11} a and b we show histograms of $\delta_0^2$, the 
mean-square distance to the lowest energy state, near $T=T_f$. 
Both histograms exhibit a multi-peak structure, with a narrow peak
at low $\delta_0^2$ that corresponds to the native state. The
thermodynamic weight of the native state is, by definition,
approximately 50\% in both cases. However, the distributions    
differ in shape, which is important from the viewpoint of kinetics. 
From Figs.~\ref{fig:11} a and b one concludes that the $\delta_0^2$ 
distribution is much more rugged for sequence 6 than for sequence 4. 
This suggests that, at $T=T_f$, folding is fastest for sequence 4, 
the sequence with highest $T_f$. We also explored using $Q$ rather 
than $\delta_0^2$ as reaction coordinate, with similar results.     

The $Q$ distribution at $T_f$ has been studied previously
by Ref.~\cite{Sali:94}, using a lattice model. 
Here the distributions for a folding and a 
non-folding sequence were compared, and these were found  
to be almost identical. This may seem to contradict our 
findings, and suggest that there is a difference between 
the two models. However, it should be remembered that Ref.~\cite{Sali:94} 
did not study the full $Q$ distribution, but rather the $Q$ 
distribution corresponding to maximally compact structures only.     
  
In Figs.~\ref{fig:11} c and d we show the probability distributions 
of $E$ and $\rgyr^2$ at $T\approx T_f$ for sequence 4. 
Also shown are the contributions
to these distributions from conformations with $\delta_0^2<0.04$
and $\delta_0^2>0.04$, respectively. This corresponds to a simplified
two-state picture where each conformation is classified as either
native-like or not. The shape of the $\delta_0^2$ distribution shows 
that such a classification is feasible in an essentially unambiguous way.
 
As one might have expected from the sharpness of the
peak in the specific heat, Fig.~\ref{fig:11} c shows that these two ``states'' 
differ significantly in energy, although the total $E$ distribution 
is unimodal. However, the average size of the molecule is very similar for the 
two states. This fact clearly demonstrates that the compactification occurs prior to the 
dominance of the native state setting in.     

These results may be compared with those of Ref.~\cite{Socci:95}, 
where a detailed study of the behavior at $T_f$ was 
carried out for a lattice model. Here the total number of contacts, $C$, 
rather than $\rgyr^2$ was used as a measure of compactness.
In contrast to the results shown in Figs.~\ref{fig:11} c and d,  
the probability distributions of both $E$ and $C$ were found 
to be bimodal for a sequence with high $T_f$. This may reflect a
genuine difference between lattice and off-lattice models. 

\section{Summary}

We have extended the off-lattice protein model of Ref.~\cite{Irback:96b} to 
three dimensions. In doing so, one key issue has been  the choice of species-independent 
local interactions that balance the species-dependent non-bonded Lennard-Jones 
interactions. The local bond and torsional 
angle interactions were chosen to satisfy two criteria:
\begin{itemize}
\item The resulting low temperature configurations should at least 
qualitatively reproduce the local angle distributions and correlations 
of functional proteins.
\item It should be possible to produce good folders; i.e. there should be sequences
with thermodynamically stable structures at not too low temperatures.
\end{itemize}
It turns out that the presence of local interactions is necessary for satisfying 
these criteria.  Among the two local interactions, the bond angle one is 
the most crucial one in this respect.

After having specified the interaction to meet these requirements we study the
thermodynamic behavior for six different sequences. The following generic 
behavior emerges:
\begin{itemize}
\item As the temperature is decreased, a gradual compactification takes place. 
\item  In the compact state a sequence dependent folding transition occurs, 
where the good folding sequences are characterized by  a higher folding 
temperature. In terms of the specific heat,  these good folders also have 
more pronounced peaks.
\item  In the state from which the transition to the native state occurs, a
large fraction of the native contacts are already formed. The contacts
missing are mainly those corresponding to large topological distance along the
chain.
\end{itemize}
This picture is consistent with what is observed for lattice 
models~\cite{Karplus:95,Socci:95}. 
A few minor, but significant, differences should be mentioned though.
\begin{itemize}
\item 
We do not observe bimodal distributions of energy or compactness 
as in Ref.~\cite{Socci:95}. In a simplified two-state picture, 
the two coexisting states at the folding temperature do correspond to different energy
distributions, but the overlap between the two is large enough to blur this
effect. 
\item For the sequences studied, high folding temperature is accompanied by 
a relatively low temperature for the peak of the specific heat. This is in 
contrast to what was reported in Ref.~\cite{Socci:95}.
\end{itemize}
Finally, we stress the fact that we have studied sequences with only two 
types of residues. In Refs.~\cite{Shakhnovich:94,Yue:95} a number of 
binary (two-letter code) sequences were studied in a 3D lattice model. 
The two-letter code was found to be insufficient in the sense that these 
sequences did not have unique native structures. In our model, many  
binary sequences do have unique native structures.
                     
Our ability to map out the thermodynamics of the 3D off-lattice model, 
relies heavily upon the efficiency of the dynamical-parameter 
algorithm of Refs.~\cite{Irback:95b,Irback:96b}. As it stands, the results for each 
$N=20$ chain require 70 CPU hours on an Alpha DecStation 200. We feel 
confident that additional algorithmic efficiency improvements can be made,  
which will enable us to probe longer and more 3D chains than reported in 
this work. 

\newpage

\newcommand  {\Biopol}   {{\it Biopolymers\ }}
\newcommand  {\BC}       {{\it Biophys.\ Chem.\ }}
\newcommand  {\BJ}       {{\it Biophys.\ J.\ }}
\newcommand  {\COSB}     {{\it Curr.\ Opin.\ Struct.\ Biol.\ }}
\newcommand  {\EL}       {{\it Europhys.\ Lett.\ }}
\newcommand  {\JCC}      {{\it J.\ Comp.\ Chem.\ }}
\newcommand  {\JCoP}     {{\it J.\ Comp.\ Phys.\ }}
\newcommand  {\JCP}      {{\it J.\ Chem.\ Phys.\ }}
\newcommand  {\JMB}      {{\it J.\ Mol.\ Biol.\ }}
\newcommand  {\JP}       {{\it J.\ Phys.\ }}
\newcommand  {\JPC}      {{\it J.\ Phys.\ Chem.\ }}
\newcommand  {\JSP}      {{\it J.\ Stat.\ Phys.\ }}
\newcommand  {\M}        {{\it Macromolecules\ }}
\newcommand  {\MC}       {{\it Makromol.\ Chem.,\ Theory Simul.\ }}
\newcommand  {\MP}       {{\it Molec.\ Phys.\ }}
\newcommand  {\Nat}      {{\it Nature}}
\newcommand  {\NP}       {{\it Nucl.\ Phys.}}
\newcommand  {\Pro}      {{\it Proteins:\ Struct.\ Funct.\ Genet.\ }}
\newcommand  {\ProSci}   {{\it Protein\ Sci.\ }}
\newcommand  {\Pa}       {{\it Physica\ }}
\newcommand  {\PL}       {{\it Phys.\ Lett.\ }}
\newcommand  {\PNAS}     {{\it Proc.\ Natl.\ Acad.\ Sci.\ USA\ }}
\newcommand  {\PR}       {{\it Phys.\ Rev.\ }}
\newcommand  {\PRL}      {{\it Phys.\ Rev.\ Lett.\ }}
\newcommand  {\Sci}      {{\it Science\ }}
\newcommand  {\ZP}       {{\it Z.\ Physik\ }}

\newpage

\end{document}